\documentclass[10pt, a4paper]{article}
\usepackage[left=2cm, right=2cm, top=2cm, bottom=2cm]{geometry}
\usepackage[natbib=true, style=ieee, citestyle=numeric-comp, sorting=none]{biblatex}
\usepackage{booktabs, diagbox, authblk, orcidlink, subcaption}
\newcolumntype{C}[1]{>{\centering\let\newline\\\arraybackslash\hspace{0pt}}p{#1}}
\usepackage{lineno, setspace}

\begin{document}
\title{Public Goods Games in Disease Evolution and Spread}

\author{Christo Morison$^{\orcidlink{0000-0002-9350-7833}1\dagger}$, Ma\l{}gorzata Fic$^{\orcidlink{0000-0002-8089-7887}2, 3\dagger}$, Thomas Marcou$^{\orcidlink{0000-0002-9353-0566}4}$, Javad~Mohamadichamgavi$^{\orcidlink{0000-0001-7208-5996}5}$, Javier Redondo Ant\'{o}n$^{\orcidlink{0009-0009-7991-4369}6}$, Golsa Sayyar$^{\orcidlink{0000-0002-8143-3187}7}$, Alexander Stein$^{\orcidlink{0000-0003-0520-0063}8}$, Frank Bastian$^{\orcidlink{0000-0003-1910-024X}9}$, Hana Krakovsk\'{a}$^{\orcidlink{0000-0002-0062-3377}10,11}$, Nandakishor Krishnan$^{\orcidlink{0000-0002-3474-7546}12}$, Diogo L. Pires$^{\orcidlink{0000-0002-6069-7474}13}$,  Mohammadreza Satouri$^{\orcidlink{0000-0003-2397-1785}14}$, Frederik J. Thomsen$^{\orcidlink{0000-0002-3477-3883}15}$, Kausutua Tjikundi$^{\orcidlink{0000-0001-9669-8789}6,16}$ \& Wajid Ali$^{\orcidlink{0000-0001-5533-1315}17 *}$}

\affil{
{\footnotesize
$^{1}$School of Mathematical Sciences, Queen Mary University of London, London, UK.

$^{2}$Department of Theoretical Biology, Max Planck Institute for Evolutionary Biology, Pl\"on, Germany.

$^{3}$Center for Computational and Theoretical Biology, University of W\"{u}rzburg, W\"{u}rzburg, Germany.

$^{4}$Centre for Mathematical Biology, University of South Bohemia, \v{C}esk\'{e} Bud\v{e}jovice, Czech Republic.

$^{5}$Institute of Applied Mathematics and Mechanics, University of Warsaw, Warsaw, Poland.

$^{6}$Department of Computer Science, University of Turin, Turin, Italy.

$^{7}$Bolyai Institute, University of Szeged, Szeged, Hungary.

$^{8}$Evolutionary Dynamics Group, Centre for Cancer Genomics and Computational Biology, Barts Cancer Institute, Queen Mary University of London, London, UK.

$^{9}$School of Mathematical Sciences, University College Cork, Cork, Ireland.

$^{10}$Section for Science of Complex Systems, Medical University of Vienna, Vienna, Austria.

$^{11}$Complexity Science Hub Vienna, Vienna, Austria.

$^{12}$HUN-REN Centre for Ecological Research, Institute of Evolution, Budapest, Hungary.

$^{13}$Department of Mathematics, City, University of London, London, UK.

$^{14}$Institute for Health Systems Science, Faculty of Technology, Policy and Management, Delft University of Technology, Delft, Netherlands.

$^{15}$Institute of Applied Mathematics, Delft University of Technology, Delft, Netherlands.

$^{16}$ISI Foundation, Turin, Italy.

$^{17}$Department of Mathematical Sciences, University of Liverpool, Liverpool, UK.\\

$^\dagger$These authors contributed equally to this work.\\
$^*$Corresponding author: \texttt{wajid.ali@liverpool.ac.uk}}
}

\maketitle

\begin{abstract}
\noindent Cooperation arises in nature at every scale, from within cells to entire ecosystems. In the framework of evolutionary game theory, public goods games (PGGs) are used to analyse scenarios where individuals can cooperate or defect, and can predict when and how these behaviours emerge. However, too few examples motivate the transferal of knowledge from one application of PGGs to another. Here, we focus on PGGs arising in disease modelling of cancer evolution and the spread of infectious diseases. We use these two systems as case studies for the development of the theory and applications of PGGs, which we succinctly review and compare. We also posit that applications of evolutionary game theory to decision-making in cancer, such as interactions between a clinician and a tumour, can learn from the PGGs studied in epidemiology, where cooperative behaviours such as quarantine and vaccination compliance have been more thoroughly investigated. Furthermore, instances of cellular-level cooperation observed in cancers point to a corresponding area of potential interest for modellers of other diseases, be they viral, bacterial or otherwise. We aim to demonstrate the breadth of applicability of PGGs in disease modelling while providing a starting point for those interested in quantifying cooperation arising in healthcare.
\end{abstract}


\section{Introduction}

Interactions between individuals amidst an ever-changing environment provide nature with immense complexity. Modelling essential features of evolution, such as selection for advantageous traits, can in part be reduced to interrelations between entities---which can range in scale from subcellular molecules to entire organisms belonging to the same or different species. Evolutionary game theory (EGT) inspects interactions within a population, translating the payoffs of game theory into evolutionary fitness~\cite{smith1973logic}. In this framework, players are not overtly rational, and strategies (types) are inherited according to principles of Darwinian evolution rather than rationally chosen. Consequently, natural selection leads to changes in the frequency of strategies depending on their relative fitness~\cite{sigmund1999evolutionary}. Some of these strategies, at first glance, seem to contradict Darwinian selection: for instance, the emergence of behaviours favouring the group over the individual~\cite{smith1974theory}.

Despite natural selection being centred on competition, cooperative behaviour and relationships arise across nature. Symbiosis can take many forms, such as services like protection (e.g., plant-ant~\cite{calixto2018protection}) or reproductive services (e.g., pollination~\cite{waser2014pollination} or seed dispersal~\cite{iluz2011zoochory}), often in exchange for resources like food or nutrients~\cite{biedermann2020ecology, nepi2018nectar, stadler2005ecology}. It can also be found at the microscale: for instance, eukaryotes evolved from primitive unicellular organisms (including the predecessors to mitochondria), once capable of independent existence, via a cooperative process called endosymbiosis~\cite{sagan1967origin, zachar2018farming, lopez2020syntrophy}. Because cooperation is an overarching theme across the scales of organisation, EGT provides a methodological path towards a deeper understanding of evolutionary processes.

While cooperation may arise via many different mechanisms~\cite{nowak2006five}, the issues surrounding allocations of resources and distributions of costs are common. This is aptly described by public goods games (PGGs), where individuals can contribute to a public good, which they then benefit from, regardless of whether they contributed or not~\cite{fehr2002altruistic}. The two-player version of a PGG is perhaps the most well-known game: the Prisoner's Dilemma (PD), introduced in an experiment by Dreshner and Flood, named by Tucker~\cite{straffin1980prisoner, tucker1983mathematics} and used to study cooperation for decades~\cite{axelrod1981evolution}. Both of these will be more formally introduced in the following section, as this paper surveys their use in a prominent area of mathematical biology: disease modelling.

Two significant applications of mathematical modelling in healthcare are the spread of pathogens and cancer evolution. Both of these have been widely described with a variety of methods. For instance, population dynamics can be explored with differential equations~\cite{wodarz2014dynamics, brauer2019mathematical}, agent-based models~\cite{frias2011agent, bravo2020hybrid}, or stochastic processes~\cite{allen2008introduction, durrett2015branching}, whereas including social structure involves borrowing tools from network science~\cite{loscalzo2017network}. Notably, PGGs can be applied in many contexts, albeit in different ways: the social aspects of epidemiology lend themselves to the emergence of cooperative behaviour via implementations such as quarantine or vaccination mandates~\cite{alam2020based}. On the other hand, cells that evolve to be cancerous are both defecting from the healthy population~\cite{cleveland2012physics} and cooperating with one another in support of the new entity called the tumour~\cite{coggan2022role}, whose cells are sometimes even considered a distinct species~\cite{turajlic2019resolving, pienta2020convergent, brown2023updating}. In reviewing these two areas, we will discuss the appearance of PGGs and discuss some ways in which each domain may be able to inform the other.

\section{Public Goods Games}

\begin{figure}[ht]
    \centering
    \includegraphics[width = 0.8\textwidth]{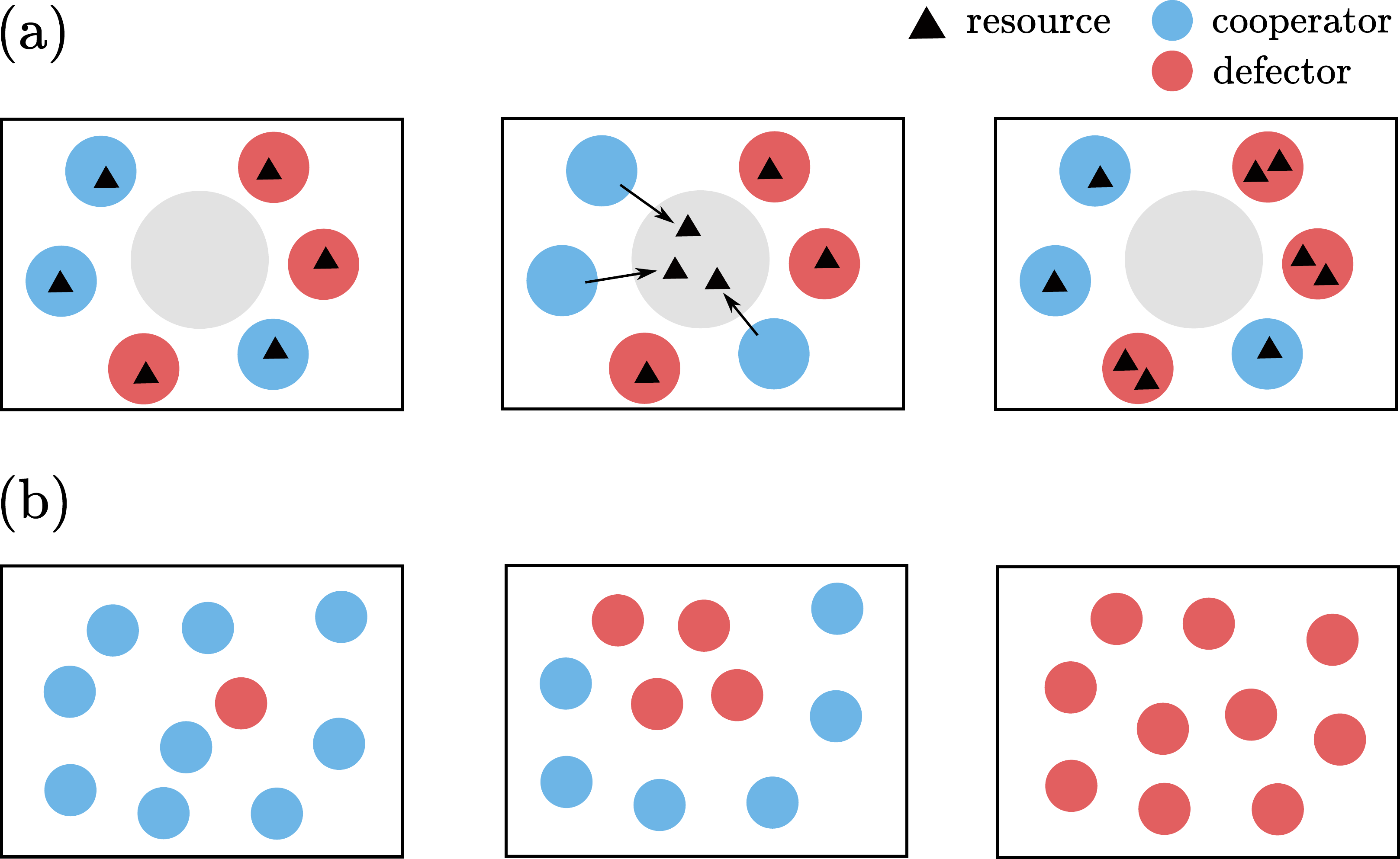}
    \caption{\textbf{(a)}~In public good games, individuals are presented with a choice to either contribute an amount $c$ to a common pot (Cooperate) or not (Defect). All contributions are then multiplied by a factor $r$ (in the example depicted, $r = 2$) and shared equally amongst all players, regardless of their contribution. Hence, a temptation to free ride arises, as defectors still benefit from the public good without incurring the contribution cost. \textbf{(b)}~Cooperators sustain a cost to supply the whole population with the public good; however, defectors can take advantage of this public good and not suffer said cost. Thus, defectors have an evolutionary advantage, as their payoff ($P_d$) is always higher than the cooperators' payoff ($P_c < P_d$). Subsequently, a defector introduced in a population of cooperators would be favoured by selection and reproduce faster, eventually leading to their fixation in the population.}
    \label{fig:pgg}
\end{figure}

In the context of EGT, PGGs are used to study how cooperative strategies arise or collapse over time. The dynamics that emerge showcase the tension between the benefit of the group and the self-interests of individuals~\cite{sigmund2010calculus}. In this game, we consider a group of $N$ players, each endowed with a resource $c$, which they can invest in a public pool (a strategy called Cooperate) or not (a strategy called Defect). The total investment within the group is then multiplied by a factor $r$, where $1<r<N$, and distributed back to all players, regardless of individual contribution (see Figure~\ref{fig:pgg}a). If $n_c$ players cooperate, the payoff of a defector is $P_d = rc n_c / N$, and that of a cooperator is $P_c = P_d - c$ since the cooperators incur the additional cost of investing in the public good~\cite{hauert2004dynamics}. Note that in many biological processes, the benefit of growth factors is not linear (that is, $r$ is not a constant in $n_c / N$). Nevertheless, though there are incentives for players to invest for the benefit of the group, there also exist incentives to free ride off of others' contributions~\cite{sigmund2010calculus, perc2013evolutionary}  (see Figure~\ref{fig:pgg}b). A rise in the frequency of defectors may lead to a phenomenon commonly referred to as the ``tragedy of the commons'', where selfish behaviour leads to a depletion of the common good~\cite{hardin1968tragedy}.

When $N = 2$, the PGG can be formulated as the PD~\cite{hauert2003prisoner, axelrod1981evolution, doebeli2005models, zhang2011eco}. Similarly to the multiplayer game, cooperation involves costs $c$ (equivalent to the net cost $c-rc/N$ in the PGG) and brings a benefit $b$ (equivalent to $rc / N$ in the PGG), which are symmetric to both players. Assuming that $b > c > 0$, this game can be summarised by the matrix representation in Table~\ref{table:PD}.  

\begin{table}[ht]
    \centering
\begin{tabular}{c|c c}
\toprule
  \diagbox{Player 1}{Player 2} & Cooperate & Defect\\ \midrule
     Cooperate & $b - c$ & $-c$\\
     Defect & $b$ & $0$ \\ \bottomrule
\end{tabular}  
    \caption{The matrix representing the payoffs received by Player 1 with respect to its strategies and the strategies of Player 2 in the Prisoner's Dilemma.}
    \label{table:PD}
\end{table}

Historically, the emergence of cooperation was examined with well-mixed population models that assume all-to-all interactions~\cite{hofbauer1998evolutionary}. Nevertheless, these models fail to capture the complexity of real-life interactions~\cite{vega2007complex}. Introducing networks into the framework of EGT allows for the inclusion of some of those complexities, like spatial and temporal structure or social relations. Individuals are then represented as nodes and interactions as edges. Players interact locally with their neighbours, either with one at a time, in two-player games, or with all neighbours simultaneously, in $N$-player games. Strategies are updated according to a specified rule (of which there are many~\cite{masuda2009directionality}), allowing for successful ones to spread.

By studying various network structures, it has been shown that the interaction topology impacts strategy evolution~\cite{roca2009evolutionary}. For example, the spread of cooperation in the PD varies across lattice~\cite{szabo1998evolutionary, nowak1992evolutionary}, small-world~\cite{abramson2001social}, regular~\cite{santos2005scale}, and real-world networks~\cite{lozano2008mesoscopic}. The impact of other topological aspects of networks on cooperation has been studied: some examples include the average degree of the network~\cite{tang2006effects}, the degree distribution heterogeneity~\cite{santos2006evolutionary}, the presence of hubs on scale-free networks~\cite{santos2006graph}, as well as the strategic placement of cooperators~\cite{yang2021strategically}. Likewise, when considering PGGs with more than two players, introducing an interaction structure influences the emergence of cooperation. This holds without considering additional features and when incorporating mechanisms such as punishment strategies, reputation, voluntary participation, or social diversity. This has been demonstrated, for instance, on lattice networks~\cite{hauert2002volunteering, brandt2003punishment, helbing2010}, as well as on regular graphs and heterogeneous scale-free networks~\cite{santos2008social, cao2010evolutionary}.

Additionally, multilayer networks can represent multiple types of interactions and individuals, add temporal and spatial context or depict communities~\cite{pilosof2017multilayer, de2023more}. They are shown to have a significant impact on the evolutionary dynamics and can promote the evolution of cooperation in the PD~\cite{gomez2012evolution, wang2015evolutionary}, multiplayer PGGs~\cite{wang2012probabilistic} and when several games are present~\cite{santos2014biased, deng2018multigames, wu2021evolution}.

\section{Applications}

PGG models have found applications in both epidemiology and oncology, two crucial areas of public health. Game theoretic tools have been used for decades to study infectious diseases~\cite{bauch2004vaccination} and cancer~\cite{vincent1996modeling, tomlinson1997game}. Spurred by the recent COVID-19 pandemic and high incidence rates of cancer worldwide, they have been used even more widely in these areas. Here, we focus on the appearance of PGGs in both contexts by reviewing the literature and suggesting ways the two fields may learn from one another.

\subsection{Epidemics}

Epidemiology is the study of the distribution and determinants of health-related events in populations~\cite{giesecke2017modern}. With the help of mathematical models, one can monitor the occurrence of infectious diseases and the course of epidemics, help design public health responses and plan for future incidences. Though Kermack and McKendrick's susceptible-infected-recovered (SIR) model~\cite{kermack1927contribution} is almost a century old, the first appearance of a game theoretical model in epidemiology was in 2004, when Bauch and Earn~(2004) described a vaccination game~\cite{bauch2004vaccination}. Since then, many PGGs have been used to model epidemiological phenomena, from herd immunity to antibiotic resistance, as well as non-pharmaceutical interventions like social distancing and mask mandates, as shown in Table~\ref{table:epidemics}. Importantly, social interactions can be incorporated in each case by adding structure to the populations.

\begin{table}[ht]
    \centering
    \begin{tabular}{C{6cm} C{5cm} C{4cm}}
    \toprule
        Public good & Cooperate & Defect\\ \midrule
        Herd immunity (e.g.~\cite{dees2018public}) & Immunisation 
        & Susceptibility\\
        Pathogen-free environment (e.g.~\cite{yong2021noncompliance})& Following non-pharmaceutical interventions, such as mask mandates or social distancing & Not following non-pharmaceutical interventions\\
        Sensitivity to antibiotics (e.g.~\cite{porco2012does})& Not overusing antibiotics & Overusing antibiotics\\ \bottomrule
    \end{tabular}  
    \caption{Examples of public goods found in epidemiological modelling, with corresponding cooperating and defecting strategies.}
    \label{table:epidemics}
\end{table}

\begin{figure}
    \centering
    \includegraphics[width = \textwidth]{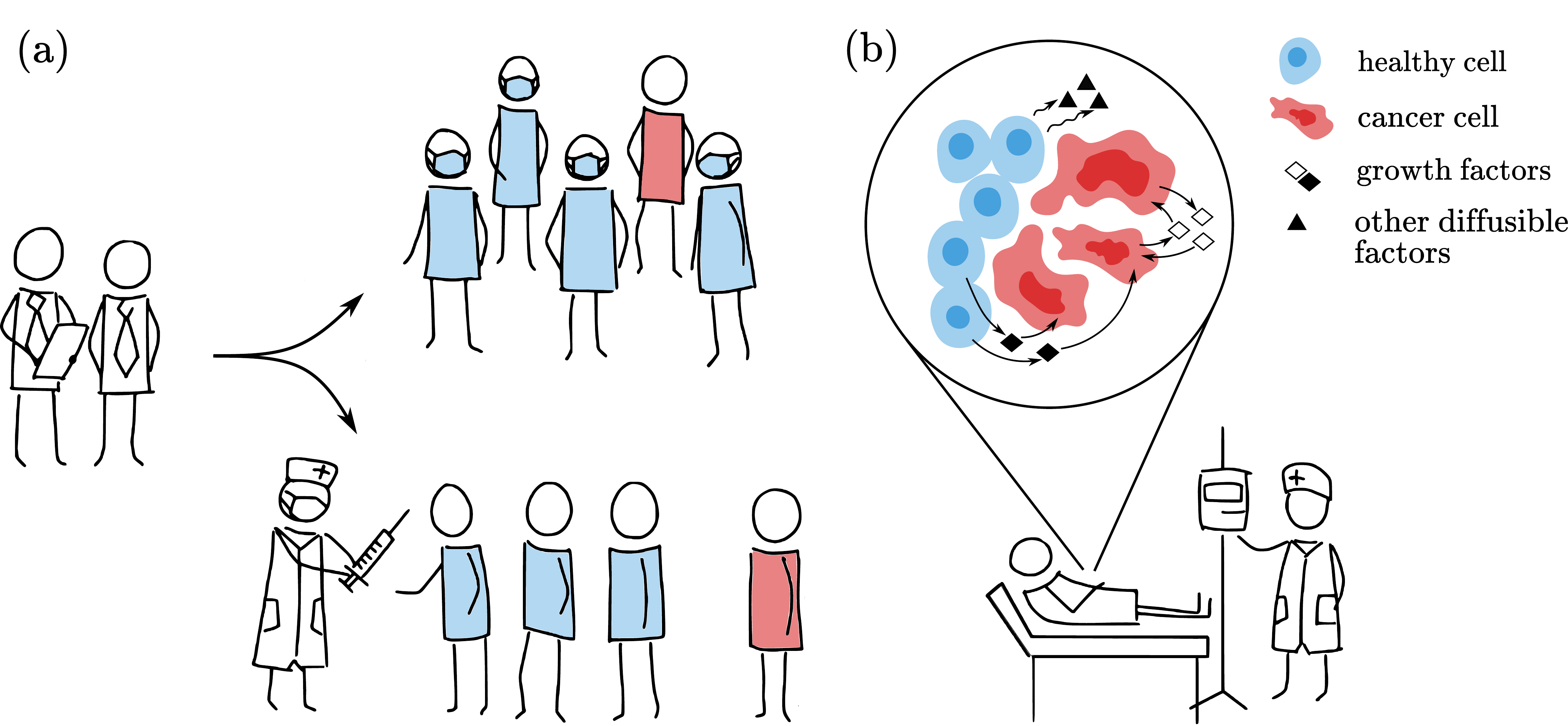}
    \caption{\textbf{(a)} During an epidemic, policy-makers are faced with implementing intervention policies aiming at disease containment and eradication. These approaches include vaccines as well as non-pharmaceutical interventions such as mask mandates, lockdowns, etc. PGG models can inform policy-makers by providing insights into how both infectious disease and individual behaviours evolve in the population. Here, defecting individuals are shown in red and cooperating individuals in blue. \textbf{(b)} Within the complex tumour microenvironment, healthy and cancerous cells are constantly exchanging information and resources. While tumour cells may free ride by not producing some diffusible factors, they can also cooperate amongst themselves. Clinicians can use evolutionary insights from these underlying dynamics to inform their therapeutic protocols.}
    \label{fig:epicanc}
\end{figure}

Herd immunity is established within a population when a sufficiently large fraction has undergone immunisation, either naturally or via vaccination, ensuring that the disease cannot persist as an endemic condition~\cite{brauer2019mathematical}. Because it safeguards individuals against the onset of infectious diseases and is both non-excludable and non-rivalrous, herd immunity can be conceptualised as a public good~\cite{dees2018public}. Through this lens, immune individuals are cooperators, and those susceptible are defectors. If enough of the population is immune---that is, if $n_c / N$ is greater than some threshold, typically around 90\%~\cite{dees2018public}---the potential drawbacks linked to getting vaccinated may surpass the risks posed by the actual infection. As a result, a free riding strategy may be favoured by some individuals (see Figure~\ref{fig:epicanc}a). When sufficiently many contributors are necessary for the public good to be reaped, such as in the vaccination game with herd immunity, it is called a threshold PGG~\cite{giubilini2021vaccination}. The game can be introduced into classical SIR-type models to better understand individual behaviour in the face of voluntary vaccination~\cite{alam2020based}. 
These models can be further enriched by introducing an incubation period between infection and symptom onset. Soltanolkottabi \textit{et al.}~(2020) show that the inclusion of this time delay can fundamentally change the epidemic dynamics, leading to fewer vaccinated and free riding individuals and more infections~\cite{soltanolkottabi2020game}. The model also includes a community structure wherein individuals get vaccinated when their (vaccinated) neighbours obtain a higher payoff. Fu \textit{et al.}~(2011) relax that assumption, introducing uncertainty into the decision to vaccinate. The model showcases social structure's ability to either promote voluntary vaccinations or facilitate disease spread~\cite{fu2011imitation}. Wang \textit{et al.}~(2020) further investigate the motivation behind vaccination decisions. Two reasons are considered for getting vaccinated: conforming and increasing one's payoff. A multilayer network approach then decouples the epidemic dynamics from the vaccination behaviour and captures the multi-levelled nature of human interactions~\cite{wang2020vaccination}. These results display the importance of individual motivation and social structure in vaccination campaigns.

PGGs have also been used to model other dilemmas arising in the wake of epidemics, such as wearing masks~\cite{tori2022study, kabir2021prosocial} and quarantining~\cite{amaral2021epidemiological}. Here, rather than herd immunity, the pathogen-free environment can be seen as a public good~\cite{yong2021noncompliance}. For example, Traulsen \textit{et al.}~(2023) highlight the importance of individual preference in designing intervention campaigns, including mask mandates, social distancing and vaccinations~\cite{traulsen2023individual}. Hence, real-world dynamics depend greatly on players' perceptions of payoffs and risks associated with becoming infected or vaccinating. It is therefore crucial for accurate modelling to consider payoff calibration depending on players' properties like demographics, location, frequency of interactions and attitude towards vaccination, as well as properties of the disease itself~\cite{piraveenan2021optimal, chapman2012using, betsch2013inviting}.

Although the whole population can enjoy public goods, sometimes the underlying PGGs are played only by a fraction of the individuals. In the case of antibiotic resistance evolution, those individuals are clinicians, and the public good in question is the sensitivity to antibiotics~\cite{porco2012does}. From the point of view of a doctor, it is always better to treat their patient with antibiotics, even when the diagnosis is uncertain. However, drug overuse accelerates the evolution of resistance, leading to the tragedy of the commons~\cite{porco2012does}. Thus, PGGs can model the use of antibiotics and provide insight into possible strategies to avoid antibiotic overuse and resistance evolution~\cite{chen2018social, colman2019medical, diamant2021game}.

\subsection{Cancer}

Cancers are evolutionary diseases initiated in a process termed carcinogenesis, in which normal cells transform into malignant tumour cells~\cite{nowell1976clonal}. EGT models have been used to study interactions occurring during disease progression and treatment at different scales: between cancer cells and cancer cells~\cite{lee2023clonal}, cancer cells with the tumour microenvironment~\cite{buhler2021mechanisms, bukkuri2022glut1}, and cancer cells with the physician~\cite{stankova2019optimizing}. In particular, frequency-dependent modelling in oncology began with tools from optimal control~\cite{martin1992optimal} before being formalised as EGT~\cite{vincent1996modeling, tomlinson1997game}. One of its most significant applications has been modelling the emergence of resistance to treatment~\cite{coggan2022role, wolfl2022contribution}. In particular, PGGs have been used to study the evolution of cooperation in cancer, both from the perspective of cancer cells defecting from the healthy population as well as cooperation amongst the cancer cells themselves~\cite{archetti2019cooperation}, as depicted in Table~\ref{table:cancer}.

\begin{table}[ht]
    \centering
    \begin{tabular}{C{6cm} C{5cm} C{3cm}}
    \toprule
        Public good & Cooperate & Defect\\ \midrule
        Growth factors (e.g.~\cite{archetti2015heterogeneity}) & Producing growth factors (cancerous or healthy cells)& Not producing (only cancer cells)\\
        Other diffusible factors, such as those promoting neoangiogenesis or disabling an immune response (e.g.~\cite{axelrod2006evolution}) & Producing diffusible factors & Not producing\\
        Adhesion (e.g.~\cite{archetti2019cooperation}) & Producing adhesion molecules & Not producing\\\bottomrule
    \end{tabular}  
    \caption{Examples of public goods found in cancer modelling, with corresponding cooperating and defecting cell types.}
    \label{table:cancer}
\end{table}

Carcinogenesis can be viewed as cancer cells free riding on a homeostatic (i.e.~under regulation to maintain stability), healthy population; this has led to cancer cells being sometimes called ``cheater cells''~\cite{cleveland2012physics}. On the other hand, an established tumour can also be modelled as a collection of cooperating subclones~\cite{cleary2014tumour}. Many cancer processes rely on the production of diffusible signalling factors by the cancer cells to promote growth~\cite{hanahan2000hallmarks} (see Figure~\ref{fig:epicanc}b). However, producing these factors comes with a cost, such that it is often beneficial for an individual to free ride on the resources produced by others. Here, the benefit of growth factors is often nonlinear, modelled as sigmoidal in concentration~\cite{cornish2013fundamentals}. For example, Archetti \textit{et al.}~(2015) modelled insulin-like growth factor II (IGF-II) as a public good amongst pancreatic cancer cells in mice. This led to coexistence between cooperators (producers of IGF-II) and defectors (non-producers), whose bistability was predicted by PGG dynamics~\cite{archetti2015heterogeneity}.

Beyond growth factors shared amongst cancer cells, other diffusible factors such as those promoting neoangiogenesis (the production of new blood vessels to supply the tumour with resources such as oxygen) or those disabling the immune system~\cite{joyce2009microenvironmental} can be considered public goods. Axelrod \textit{et al.}~(2006) argue that this kind of cellular-level cooperation between partially transformed tumour cells is possible because several hallmarks of cancer---sustaining proliferative signalling, inducing angiogenesis, and activating tissue invasion and metastasis~\cite{hanahan2000hallmarks}---involve shareable resources~\cite{axelrod2006evolution}. Threshold PGGs can also model direct contact between cells and their neighbours. Here, the public good (adhesion between cells) is provided so long as a minimum number of cells contribute adhesion molecules~\cite{archetti2019cooperation}.

Moreover, Archetti~(2021) has proposed leveraging cooperation amongst cancer cells to improve treatment. Genetically engineering some cancer cells to free ride (by not producing a certain growth factor) led to a decrease in the proliferation rate of the tumour population~\cite{archetti2021collapse}. Hence, PGGs could join other evolutionary concepts in mathematical oncology that have led to novel treatment strategies such as adaptive~\cite{gatenby2009adaptive, zhang2017integrating}, double-bind~\cite{basanta2012exploiting} and extinction therapies~\cite{gatenby2019first, gatenby2020integrating}.

\section{Discussion}

Mathematical models in epidemiology and oncology can inform public health officials and clinicians in their quest to manage and eradicate infectious diseases and cancers. The evolution of tumours and the spread of pathogens have been studied with PGGs; however, their uses have been fairly disjoint. Here, we outline some opportunities for future avenues of research.

The evolution of resistance is a central issue in both fields: cancers not eradicated with a first line of therapy are prone to decreasing treatment sensitivity~\cite{gatenby2009change}; likewise, antibiotic resistance threatens one's ability to eliminate pathogenic bacteria~\cite{levy1998challenge}. Just as sensitivity to antibiotics is a public good to a population of (possibly infected) individuals (see Table~\ref{table:epidemics}), therapeutic sensitivity can be thought of as a public good within a cancer cell population. In both cases, resistance is positively selected for when treatment is applied. On the other hand, it has been shown that bacteria resistant to antibiotics can help shield their sensitive counterparts, increasing the survival capability of the whole population~\cite{lee2010bacterial}. With these parallels in mind, recall Archetti's engineered defector cells and their role in collapsing intratumour cooperation~\cite{archetti2021collapse}. Such a defecting population will spread in the cancer population under clonal selection. If this subclone could be kept sensitive to a certain treatment, then once it subsumes the cooperating population, the tumour might have a better chance of being eradicated by said treatment. Similarly, in epidemiology, one might create a strain of bacteria that doesn't produce a certain metabolite~\cite{lenski2001games}, thus defecting in the PGG though not evolving antibiotic resistance.

Introducing network structure in EGT models allows for more accurate representations of real-world dynamics~\cite{szabo2007evolutionary}. For instance, knowledge of the underlying social network during an epidemic can help identify the most influential individuals to vaccinate~\cite{christley2005infection}. However, during dynamical processes---such as the evolution and spread of diseases---these structures are rarely static~\cite{perc2010coevolutionary} and often co-evolve with the strategies~\cite{pacheco2006coevolution, zimmermann2004coevolution} or even independently~\cite{perc2008restricted,kun2009evolution}. Another class of spatial models considers individuals moving through networks, allowing for the inclusion of complex, ever-changing social interaction patterns~\cite{broom2021review}. As such, tools from evolving network theory can sharpen current EGT models. Much like in mathematical epidemiology, the inclusion of spatial structure in cancer models impacts the evolution of cooperation~\cite{altrock2015mathematics, archetti2016cooperation, renton2021cooperative} (as reviewed in~\cite{coggan2022role}) and could also benefit from dynamical networks.

Another recent addition to EGT that shows promise to improve treatment modelling are Stackelberg games, which can describe the leader-follower dynamics between a clinician and a tumour. Here, the cancer cells are themselves also playing an evolutionary game, which allows the clinician to apply evolutionarily-informed treatments by anticipating the evolution of the disease~\cite{stein2023stackelberg, kleshnina2023game}. This two-layer framework may find parallels with epidemiological modelling: policy-makers involved in infectious disease response can be considered the leader whose knowledge of the evolving epidemic informs their decision-making. For example, targeting vaccinations can be optimised in pursuit of the public good of herd immunity~\cite{hu2023stackelberg}. Stackelberg evolutionary game theory may help formalise these optimal control problems in both cancer and epidemic modelling.

PGGs appear in many areas of mathematical biology and point to unexpected connections between disparate fields of research. Though game theoretic models are informative in understanding the key interactions within a system, real-world data is nevertheless crucial to effectively translate theoretical insights into practical applications. As mathematical modelling in healthcare matures, active discourse between theoreticians, experimenters, clinicians and policy-makers is vital to ensure appropriate model predictions and health interventions.

\subsection*{Acknowledgements}
This project has received funding from the European Union's Horizon 2020 research and innovation programme under grant agreement No.~955708. The authors would like to gratefully acknowledge Chaitanya S. Gokhale and Vlastimil K\v{r}ivan for their guidance; Esther Baena, Mark Broom, Joel Brown, Johan Dubbeldam, Rudolf Hanel, Weini Huang, Kieran Sharkey, Kate\v{r}ina Sta\v{n}kov\'a, Yannick Viossat and Benjamin Werner for their input during the preparation of the manuscript; and \'Ad\'am Kun, Jacek Mi\c{e}kisz, Daniela Paolotti, Gergely R\"ost, Maria Luisa Sapino, Sebastian Wieczorek and the other members of the EvoGamesPlus consortium for their support. CM would like to thank the Center for Computational and Theoretical Biology at the University of W\"{u}rzburg for hosting a visit to work on this manuscript.

\subsubsection*{Statement of Contribution}
Conceptualisation (all authors), Writing -- Original Draft (all authors), Writing -- Review \& Editing (all authors), Visualisation (FJT), Project Administration (WA).

\subsubsection*{Competing interests}
The authors declare no competing interests.

\newpage
\printbibliography
\end{document}